\documentclass[apjl]{emulateapj}
\usepackage{hyperref}
\usepackage{amsmath}

\shorttitle{Weak Lensing of IDCS~J1426.5+3508}
\shortauthors{W. Mo et al.}

\begin{document}

\title{IDCS~J1426.5+3508: Weak Lensing Analysis of a Massive Galaxy Cluster at $z=1.75$}

\author{
Wenli Mo\altaffilmark{1},
Anthony Gonzalez\altaffilmark{1},
M. James Jee\altaffilmark{2},
Richard Massey\altaffilmark{3},
Jason Rhodes\altaffilmark{4,5}, 
Mark Brodwin\altaffilmark{6},
Peter Eisenhardt\altaffilmark{4},
Daniel P. Marrone\altaffilmark{7},
S. A. Stanford\altaffilmark{8},
and
Gregory R. Zeimann\altaffilmark{9,10,11}
}

\altaffiltext{1}{Department of Astronomy, University of Florida, Bryant Space Science Center, Gainesville, FL 32611, USA}
\altaffiltext{2}{Department of Astronomy and Center for Galaxy Evolution Research, Yonsei University, 50 Yonsei-ro, Seoul 03722, Korea}
\altaffiltext{3}{Institute for Computational Cosmology, Durham University, South Road, Durham DH1 3LE, UK}
\altaffiltext{4}{Jet Propulsion Laboratory, California Institute of Technology, Pasadena, CA 91109, USA}
\altaffiltext{5}{California Institute of Technology, Pasadena, CA 91125, USA}
\altaffiltext{6}{Department of Physics and Astronomy, University of Missouri, 5110 Rockhill Road, Kansas City, MO 64110, USA}
\altaffiltext{7}{Steward Observatory, University of Arizona, 933 North Cherry Avenue, Tucson, AZ 85721, USA}
\altaffiltext{8}{Department of Physics, University of California, Davis, One Shields Avenue, Davis, CA 95616, USA}
\altaffiltext{9}{Department of Astronomy \& Astrophysics, The Pennsylvania State University, University Park, PA 16802, USA}
\altaffiltext{10}{Institute for Gravitation and the Cosmos, The Pennsylvania State University, University Park, PA 16802, USA}
\altaffiltext{11}{Department of Astronomy, University of Texas, Austin, TX 78712, USA}

\begin{abstract}

We present a weak lensing study of the galaxy cluster IDCS~J1426.5+3508 at $z=1.75$, which is the highest redshift strong lensing cluster known and the most distant cluster for which a weak lensing analysis has been undertaken. Using F160W, F814W, and F606W observations with the {\it Hubble Space Telescope}, we detect tangential shear at $2\sigma$ significance. Fitting a Navarro-Frenk-White mass profile to the shear with a theoretical median mass-concentration relation, we derive a mass $M_{200,\mathrm{crit}}=2.3^{+2.1}_{-1.4}\times10^{14}$~M$_{\odot}$. This mass is consistent with previous mass estimates from the Sunyaev-Zel'dovich (SZ) effect, X-ray, and strong lensing. The cluster lies on the local SZ-weak lensing mass scaling relation observed at low redshift, indicative of minimal evolution in this relation.

\end{abstract}


\section{Introduction}

Massive, high-redshift galaxy clusters, though rare, provide valuable information about cosmological parameters, structure formation, and galaxy evolution. The redshift regime $z>1.5$ represents a critical era during which significant star formation occurs in cluster galaxies \citep[e.g,][]{tran10, brodwin13, bayliss14, webb15b, alberts14}. Only a handful of clusters at redshift $z>1.5$ have been confirmed to date \citep[e.g,][]{newman14, papovich10, zeimann12, muzzin13, tozzi15, webb15}, thanks in part to large X-ray, infrared, and Sunyaev-Zel'dovich (SZ) effect surveys \citep[e.g.,][]{eisenhardt08, fassbender11, wylezalek14, bleem15}.

In this paper we focus upon the galaxy cluster IDCS~J1426.5+3508. This cluster was first discovered in the IRAC Distant Cluster Survey (IDCS). Follow-up with the {\it Hubble Space Telescope} ({\it HST}) Wide Field Camera-3 (WFC3) grism and the Low Resolution Imaging Spectrometer \citep[LRIS,][]{oke95} on the W. M. Keck Observatory spectroscopically confirms a redshift of $z=1.75$ \citep{stanford12}. With $100$~ks of {\it{Chandra}} data, \citet{brodwin15} obtain mass estimates based upon the X-ray temperature, gas mass, and the product of core-excised X-ray temperature and gas mass of $M_{500, T_{X}}=3.3^{+5.7}_{-1.2}\times10^{14}$~M$_{\odot}$, $M_{500, M_g}=2.3^{+0.7}_{-0.5}\times10^{14}$~M$_{\odot}$, and $M_{500, \mathrm{Y}_X}=2.6^{+1.5}_{-0.5}\times10^{14}$~M$_{\odot}$, respectively. In comparison, SZ observations with the Combined Array for Research in Millimeter-wave Astronomy (CARMA) indicate $M_{500, \mathrm{SZ}}=2.6\pm0.7\times10^{14}$~M$_{\odot}$ \citep{brodwin12}, implying $M_{200}=4.1\pm1.1\times10^{14}$~M$_\odot$ with a \citet{duffy08} mass-concentration relation. These mass estimates establish IDCS~J1426.5+3508 as the most massive galaxy cluster confirmed at redshift $z>1.5$. For comparison, another high-redshift cluster, XDCP~J0044.0-2033 ($z=1.58$), has mass derived from the \citet{vikhlinin09} $Y_X-$M scaling relation $M_{500,Y_X}=2.2^{+0.5}_{-0.4}\times10^{14}$~M$_{\odot}$ \citep{tozzi15}.

Moreover, IDCS~J1426.5+3508 is also the most distant strong lensing galaxy cluster. A giant gravitationally lensed arc associated with the cluster was discovered in {\it HST}/WFC3 and Advanced Camera for Surveys (ACS) imaging. An initial redshift estimate based on broadband photometry yielded $z\sim3-6$ for the arc and a lower limit of $M_{200}>2.8\times10^{14}$~M$_{\odot}$ for the cluster via a strong-lensing analysis \citep{gonzalez12}. For $\Lambda$CDM, \citet{gonzalez12} calculate that the expected number of giant arcs at this redshift and brightness is vanishingly small, highlighting the long-standing arc statistics problem \citep[see][for a recent review]{meneghetti13}. One explanation suggested to explain the observed arc in IDCS~J1426.5+3508 is that the cluster might have a substantially more concentrated density profile---and hence an enhanced lensing cross-section---than predicted for a cluster at this epoch.

Gravitational weak lensing determines the cluster mass via the distortion in shape of background galaxies due to the gravitational potential of the cluster. Unlike mass estimates derived from X-ray or SZ observations, weak lensing measures the mass distribution independent of the dynamical state or hydrostatic equilibrium of the cluster. Thus, weak lensing is also a powerful tool for calibrating X-ray or SZ mass estimates \cite[e.g.,][]{marrone12, hoekstra12, jee11}. Furthermore, the mass and concentration of the cluster can in principle be determined without a prescribed mass-concentration relationship. 

However, weak lensing is observationally challenging, particularly for distant clusters. Because the lensing cluster is at high redshift, the number of lensed background galaxies is small, decreasing the lensing signal. Also, the extent of the distortions to be measured is on the scale of the observational point spread function \citep[PSF; e.g.,][]{rhodes07, jeepsf}. Deep, space-based observations with well-understood PSFs are therefore required to obtain high fidelity measurements of the weak lensing distortion. 

In this work, we demonstrate the feasibility of weak lensing analyses at $z=1.75$. IDCS~J1426.5+3508 is the highest redshift cluster to have joint X-ray, SZ, and weak lensing observations, and thus provides an opportunity to compare whether the scaling relations between these mass estimators, derived at low redshift, remain valid at this epoch. In Section~\ref{sec:obs} we describe the observations and data analysis, including the selection of source galaxies. We outline the measurement of galaxy shapes and the correction for PSF distortion in those measurements in Section~\ref{sec:rrg}. The weak lensing shear profile is presented in Section~\ref{sec:lensing}. We discuss the mass estimation and SZ-weak lensing scaling relation in Sections~\ref{sec:mass} and \ref{sec:szwl}, respectively, and summarize our findings in Section~\ref{sec:conclusions}.

Throughout the paper, we adopt the nine-year Wilkinson Microwave Anisotropy Probe (WMAP9) cosmological parameters of $\Omega_\mathrm{M}=0.287$, $\Omega_{\Lambda}=0.713$, and $H_0 = 69.32$~km~s$^{-1}$~Mpc$^{-1}$ \citep{hinshaw13} and define $h=H_0/100$. All magnitudes are in the AB system \citep{oke83}. We take the F606W image for our shape measurement and the center of the cluster to be the location of the brightest cluster galaxy (BCG), as identified by \citet{stanford12}. Unless otherwise stated, we report masses as overdensities relative to the critical density.

\section[]{Observations}
\label{sec:obs}

For our analysis, we use {\it HST} data from Cycle 20 taken on 2012 December 19, and 2013 April 17-19, with the ACS F606W ({\it V}-band), F814W ({\it I}-band), and WFC3 F160W ({\it H}-band) filters for a total of 21760, 8108, and 4947 seconds, respectively. We also use Cycle 17 observations from 2010 July 8 and 2010 November 7 with the F814W and F160W filters for 4513 and 2612 seconds, respectively. Since the F160W data are mosaicked, the effective exposure time for the central region is 7559 seconds while the minimum exposure time for the outer regions is 1212 seconds. The cluster is also among the targets in {\it HST} GO-program 13677 (PI Perlmutter) which will add observations in WFC3 F140W, F105W, and F814W filters for future studies of this cluster.

\subsection{{\it HST} ACS Data Reduction}
We performed basic processing of the ACS data with the {\ttfamily CALACS} pipeline. Our only modification to the standard procedure was to employ the charge transfer inefficiency (CTI) correction method developed by \citet{massey10, massey14} to correct for CTI degradation prior to running {\ttfamily CALACS}. CTI extends the shapes of galaxies and, if uncorrected, can imitate the effects of weak lensing. Astrometric shifts, geometric distortion correction, sky subtraction, cosmic ray removal, and final image stacking were handled by the standard {\ttfamily TweakReg} and {\ttfamily Astrodrizzle} packages. Images were drizzled to a $0.03\arcsec$ pixel scale with a Gaussian kernel and pixel fraction of 1.0. \citet{rhodes07} showed that changing the pixel scale from the ACS native scale of $0.05\arcsec$ to $0.03\arcsec$ optimizes our ability to correct for PSF effects while the results are largely insensitive to the choice of pixel fraction. 

\subsection{Source Galaxy Catalog}
\label{sec:colorcuts}

\begin{figure}
	\centering
	\includegraphics[width=\columnwidth]{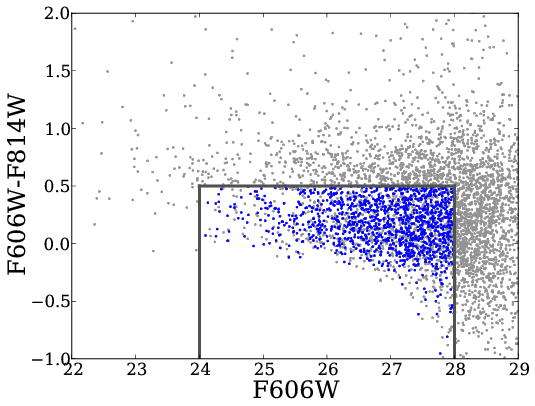}
	\caption{The color-magnitude diagram of all sources detected in the F606W image. The solid lines are the color and magnitude cuts applied to the catalog. The blue data are the sources that satisfied the magnitude, color, size, and S/N criteria for inclusion in the input catalog to our shape measurement program. Those excluded from the final catalog are shown in gray.}
	\label{fig:cmd}
\end{figure}

Starting with F606W as our detection image, we extracted a catalog of 4771 sources with {\ttfamily SExtractor} \citep{bertin96} with a criteria of $5$ connecting pixels brighter than a sky rms of $2$. We then obtained matched photometry for these sources in F814W and F160W by running {\ttfamily SExtractor} in dual image mode. Figure~\ref{fig:cmd} is the resulting color-magnitude diagram for the two bluest filters.

We identify the foreground and cluster galaxy contamination in our catalog via color selection. One benefit of employing an F606W-selected catalog is that most cluster red sequence galaxies are non-detections. Listed below are the sequence of color and size cuts applied to distinguish among foreground galaxies, cluster members, and background galaxies. We detect 3050 objects with F606W$<28.0$, the approximate $10\sigma$ depth of our observations. We first reject $295$ sources with either F606W$<24.0$ or FWHM$>0.9\arcsec$ to exclude bright and large objects that are probable foreground galaxies. Based on a \citet{bruzual03} model of a star-forming galaxy with $z_{\mathrm{form}}=6$, we next implement a color cut of F606W-F814W~$>0.5$ to eliminate $786$ additional foreground galaxies in our sample. To reject cluster members on the red sequence, we rule out $20$ objects with F814W-F160W~$>3.0$ \citep{stanford12}. We also cut $16$ objects with signal to noise $S/N<10$. 

In addition, we exclude $922$ objects based on shape measurements constraints, further discussed in Section~\ref{sec:rrg}. The final catalog for the shear analysis consists of 1011 sources, equivalent to $89$~arcmin$^{-2}$. As a cross-check, we construct a radial density profile for the remaining galaxies centered on the BCG. There is no central excess evident in the radial profile, indicating that residual contamination from blue cluster galaxies is minimal.

\subsection{Redshift Distribution}
\label{sec:redshift}

\begin{figure}
	\centering
	\includegraphics[width=\columnwidth]{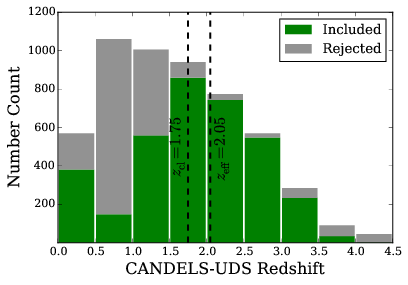}
	\caption{The redshift distribution of the galaxies in the CANDELS-UDS catalog matching the criteria $0.27\arcsec<\mathrm{FWHM}<0.90\arcsec$ and $24<\mathrm{F606W}<28$. The fraction that are included in our calculation of the effective redshift (see Section \ref{sec:redshift}) is plotted in green and the fraction rejected by our selection criteria is plotted in gray. Of the remaining CANDELS-UDS sources, 46\% are foreground contaminants with redshifts $z\leq1.75$.}
	\label{fig:zeff}
\end{figure}

Given the redshift of the cluster, compiling a clean catalog of background sources is not possible with the available photometric data. However, we can control the foreground contamination in our source catalog statistically. We estimate the redshift distribution of our source catalog using the Cosmic Assembly Near-infrared Deep Extragalactic Legacy Survey Ultra Deep Survey (CANDELS-UDS) multiwavelength catalog \citep{santini14} which provides infrared-detected sources to an F160W magnitude limit of $27.45$ in a range of filters. Employing the selection criteria described in Section~\ref{sec:colorcuts}, we found 3512 sources, 46\% of which were contaminants with $z\leq1.75$. The distribution of CANDELS-UDS redshift for sources included and excluded by our selection criteria is illustrated in Figure~\ref{fig:zeff}.

We estimate the effective redshift of the source population by calculating the lensing efficiency, the mean of the ratio of angular diameter distances,
\begin{equation}
\label{eq:beta}
    \langle\beta\rangle = \langle\mathrm{max}(0,{D_{ls}/D_{s}})\rangle,
\end{equation}
where $D_{ls}$ and $D_{s}$ are the angular diameter distances between the lens and the source and the observer and the source, respectively. We set a floor of $\beta=0$ to account for foreground sources with $z\leq1.75$. For the subset of CANDELS-UDS sources matching our selection criteria, we compute $\langle\beta\rangle=0.086$, corresponding to an effective redshift~$z_\mathrm{eff}=2.05$, and width $\langle\beta^{2}\rangle=0.017$. 

Despite CANDELS-UDS being the deepest catalog with data in our set of filters, 12\% of the galaxies in our background source catalog are fainter in F160W than the CANDELS-UDS limiting magnitude. To investigate the effects of the mismatch in detection filter threshold between our data and the CANDELS-UDS catalog, we randomly select 114 objects from a sample of $1455$ CANDELS-UDS objects near the survey limiting magnitude ($27.0<\mathrm{F160W}<27.45$) and add their redshifts to the 3512 CANDELS-UDS sources matching our selection criteria. We choose these objects because they are most likely to mimic the redshift distribution of the sources with F160W$>27.45$ missing from the CANDELS-UDS catalog. We then repeat our calculation of $\beta$ using Equation~\ref{eq:beta} with the additional 12\% of galaxies. Including the randomly selected redshift sample increases $\beta$ by $3.5\%$. This source of uncertainty for $\beta$ is a subdominant contributor to the total mass error budget.

\section{Shape Measurement}
\label{sec:rrg}

\subsection{RRG Method}

We measure galaxy distortions by applying the RRG method developed by \citet{rhodes00}, which calculates the second- and fourth-order Gaussian-weighted moments of each galaxy to determine ellipticity and correct for convolutions with the PSF. The weak lensing shear is then derived from the average ellipticity of the sources. We choose the weight function size of each object to be $w=\max(2\sqrt{ab},6)$ where $a$ and $b$ are the semi-major and -minor axes in pixels calculated from {\ttfamily SExtractor}.

For each object, the shear $\gamma$ is related to the ellipticity $e$ of that object
\begin{equation}
\label{equ:shear}
	\gamma= C\frac{e}{G},
\end{equation}
where $G=1.35$ is the shear susceptibility calculated from our data and $C=0.86^{-1}$ is a calibration factor determined from the analysis of simulated images containing a known shear \citep{leauthaud07}. The reduced shear is then
\begin{equation}
    g=\frac{\gamma}{(1-\kappa)}
\end{equation}
where $\kappa$ is the convergence. We further incorporate the width of the redshift distribution \citep{seitz97}. The observed shear $g^{\prime}$ is then
\begin{equation}
    g^{\prime} = (1+(\langle\beta^{2}\rangle/\langle\beta\rangle^2-1)\kappa)g = (1+1.55\kappa)g.
\end{equation}

In addition to the criteria presented in Section~\ref{sec:colorcuts}, we introduce more selection criteria to our background source catalog due to shape measurement limitations. The RRG method cannot correct for the distortion on objects smaller than $\sim1.5$ times the RRG weight function width of $0.18\arcsec$. Therefore, we reject $730$ galaxies with a {\ttfamily SExtractor} FWHM~$<0.27\arcsec$ for which we were not able to apply a PSF correction. Also, we exclude $192$ objects that encountered centroiding errors during shape measurement or for which the PSF correction did not converge. Therefore, our final catalog for weak lensing analysis includes 1011 sources.

\subsection{PSF Correction}

The ACS PSF is known to be both temporally and spatially variable. Thermal breathing of the telescope, dependent on the 90-minute orbit of {\it HST}, and a slow deviation in the focus on the timescale of a few weeks further complicate the PSF. In the weak lensing regime, an uncorrected PSF distortion can drown out the signal, so precise characterization is critical.

We use the publicly available TinyTim software\footnote{http://tinytim.stsci.edu/} \citep{krist97,krist11} to theoretically model the PSF distortion across our image. TinyTim simulates the PSF given a focus value and takes into account the instrument, chip number, chip position, filter, and spectrum of the object. \citet{schrabback10} notes additional variations in the PSF dependent on  relative sun angle, but finds that $97\%$ of variations can be described solely by the focus parameter. 

Our F606W image is a combination of 16 separate frames. We replicate the PSF in each frame individually by specifying a focus value determined by the {\it HST} focus model\footnote{http://focustool.stsci.edu/}, applying the average focus during the exposure period of each individual frame. 
Adopting a modified version of TinyTim\footnote{http://community.dur.ac.uk/r.j.massey/acs/PSF}, we generate a distortion-corrected grid of PSFs across the entire image with pixel scale of $0.03\arcsec$. We create the PSF model for the full stacked F606W image by combining the 16 individual PSFs weighted by the exposure time. We then interpolate the second- and fourth-order moments of the PSF model to the arbitrary positions of galaxies throughout the field of view via a third-order polynomial in $x$ and $y$.

\section{Weak Lensing Analysis}
\label{sec:lensing}

\begin{figure}
	\centering
	\includegraphics[width=1\columnwidth]{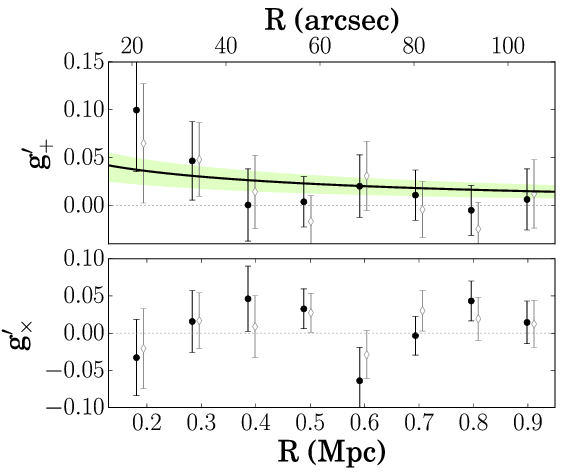}
	\caption{The reduced tangential shear profile as a function of radius from the center of the cluster is plotted as black circles in the top panel. The NFW profile using the best-fit $M_{200}$ parameter obtained with a \citet{duffy08} constraint and maximum likelihood algorithm is shown as the black solid line and the shaded region represents the $1\sigma$ confidence interval. The cross-component shear plotted as black circles in the bottom panel. The tangential shear and cross component calculated for galaxies matching the selection criteria described in Section~\ref{sec:colorcuts} and using shapes measured with the J09 method are plotted as open diamonds and show that the shapes measured with the J09 algorithm and RRG method are consistent.}
	\label{fig:shear}
\end{figure}

\begin{figure}
	\centering
	\includegraphics[width=1\columnwidth]{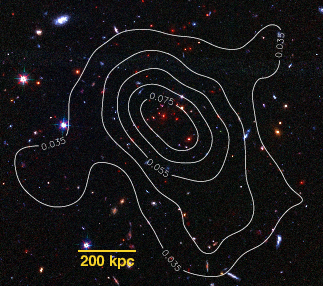}
	\caption{A combined 3-color image of galaxy cluster IDCS~J1426.5+3508 with {\it HST} F160W, F814W and F606W filters. The shear calculated with the J09 method with a relaxed galaxy selection criteria is overplotted as white contours. The centroid of the 2D reconstruction lies within $10\arcsec$ of the BCG.}
	\label{fig:obama}
\end{figure}

The weak lensing signal is expressed as the tangential component of the shear $g_+$. The cross component $g_{\times}$, defined as $g_+$ after the object is rotated by $45^{\circ}$ and represents the null statistic and noise of our measurements and should be consistent with zero in the absence of systematic errors.

We determine the shear as a function of cluster radius by radially binning our catalog objects and calculating the average $g^{\prime}_+$ and $g^{\prime}_{\times}$ of all objects in each bin. The cluster center is at the location of the BCG (RA=$14^\mathrm{h}26^\mathrm{m}32^\mathrm{s}.95$ and Dec=$+35^{\circ}08'23~6\arcsec$). The shear profile is plotted in Figure~\ref{fig:shear} where the error bars shown for the shear measurements are the statistical uncertainty of the mean shear per bin. By adding in quadrature the ratio of mean shear and uncertainty per bin for the first six bins, we calculate a significance of $2.0$ for the detection. The Pearson's correlation coefficient between the tangential and cross shear components is small ($r=-0.12$), indicating no correlation between the two components

As a test of robustness of this detection, we also employ an alternative shape measurement from the \citet[hereafter J09]{jee09} and RRG methods. Using the same galaxy selection, we find that the shapes obtained using the J09 and RRG methods yield consistent results: the typical absolute difference between the mean measured values in each bin is within $0.8\sigma$ for the tangential shear and $1.3\sigma$ for the cross component. 

The J09 approach is designed to extend to faint magnitudes. With this method, we also perform a 2-D mass reconstruction using a more relaxed set of selection criteria. Specifically, we allow a wider range in magnitudes ($22<$F606W$<29$ and $22<$F160W$<29$), a relaxed red sequence selection (F814W-F160W$<1.8$), and a decreased size selection (F606W half-light radius $r_{h}>1.2$). The centroid of the resulting 2-D mass reconstruction is close to the location of the BCG, providing further evidence that the shear is the result of the cluster weak lensing. The shear contours are overplotted on the HST image in Figure~\ref{fig:obama}. Finally, we note that the tangential shear profile using this different selection remains consistent with Figure~\ref{fig:shear} but with smaller statistical uncertainties. We continue the analysis with our original shape measurement.

\section{Mass Estimation}
\label{sec:mass}

The tangential shear is related to the density profile of the cluster assuming a Navarro-Frenk-White (NFW) profile \citep{navarro97} described by the virial radius $r_{200}$, defined as the radius inside which the mass density of the halo is equal to $200$ times the critical density, and the concentration $c=r_{200}/r_{s}$, where $r_{s}$ is the scale radius. For the radial dependence of the shear in an NFW model, we refer the reader to Equations~$14-17$ of \citet{wright00}. 

To fit the shear profile, we utilize the maximum likelihood estimation discussed in \citet{schneider00}. The log-likelihood function is defined as
\begin{equation}
\label{equ:ll}
	\ell_{\gamma} = \sum_{i=1}^{N\gamma}{\left[\frac{|e_{t,i}-g^{\prime}_{+}(\theta_{i})|^2}{\sigma^{2}[g^{\prime}_{+}(\theta_{i})]}\right]+2\ln\sigma[g^{\prime}_{+}(\theta_{i})]},
\end{equation}
where $N_{\gamma}$ is the number of galaxies with measured ellipticity, $e_{t,i}$ is the tangential ellipticity component, $\theta_{i}$ is the position of the $i$-th galaxy, and $g^{\prime}_{+}(\theta_{i})$ is the observed tangential shear at $\theta_{i}$. The dispersion of observed ellipticities approximated as $\sigma[g^{\prime}_{+}(\theta_{i})]\approx\sigma_{e}(1-|g^{\prime}_{+}(\theta_{i})|^2)$, the same as that defined in Equation~14 of \citet{schneider00}, where $\sigma_{e}=0.3$ is the intrinsic ellipticity dispersion. 

We lack the necessary signal to fit the mass and concentration simultaneously as free parameters. Thus, we derive the cluster mass with the concentration defined by the mass-concentration relation from \citet{duffy08},
\begin{equation}
	c = 5.71\Bigg(\frac{M_{200,\mathrm{crit}}}{2\times10^{12}h^{-1}M_{\odot}}\Bigg)^{-0.084}(1+z)^{-0.47},
\end{equation}
where the mass within $r_{200}$ is $M_{200,\mathrm{crit}}=200\rho_\mathrm{crit}(\frac{4}{3}\pi{r_{200}^3})$. In this case, $M_{200,\mathrm{crit}}$ is the only free parameter in the likelihood fit, and the resultant value can be applied to directly compare IDCS~J1426.5+3508 with the SZ-lensing relation defined by low redshift clusters. All objects located at radii $15\arcsec<{r}<110\arcsec$ ($130<r<950$~kpc) are included in our fit (802 sources). We exclude the strong lensing region, defined by the radius of the known strongly-lensed arc \citep{gonzalez12}, and extend to a radius that best constrains the $1\sigma$ mass confidence interval. We derive  $M_{200,\mathrm{crit}}=2.3^{+2.1}_{-1.4}\times10^{14}$~M$_{\odot}$ ($M_{500, \mathrm{crit}}=1.4^{+1.3}_{-0.9}\times10^{14}~M_{\odot}$) which corresponds to $r_{200}=0.68^{+0.16}_{-0.18}$~Mpc ($r_{500}=0.42^{+0.10}_{-0.11}$~Mpc). Changing the center to that of the X-ray observations \citep{brodwin15} yielded a mass within $1\sigma$ of the mass derived with BCG centering. The NFW profile described by our best-fit parameters is plotted in Figure~\ref{fig:shear}.

IDCS~J1426.5+3508 is the only galaxy cluster in this regime to have joint weak lensing, strong lensing, SZ, and X-ray observations. \citet{brodwin15} find good agreement between the mass estimations from X-ray, SZ, and strong lensing. Converting from $M_{500,\mathrm{crit}}$ to $M_{200,\mathrm{crit}}$ via the \citet{duffy08} relation, \citet{brodwin12} calculate an SZ mass $M_{200,\mathrm{crit}}^{\mathrm{SZ}}=4.1\pm1.1\times10^{14}$~M$_{\odot}$, \citet{brodwin15} estimate a mass from the X-ray gas mass of $M_{200}^{\mathrm{M_{gas}}}=3.8^{+1.1}_{-0.8}\times10^{14}$~M$_{\odot}$, while \citet{gonzalez12} projected a lower limit of $M_{200}>2.8^{+1.0}_{-0.4}\times10^{14}$~M$_{\odot}$ from observations of the giant arc. Our weak lensing mass of $M_{200,\mathrm{crit}}=2.3^{+2.1}_{-1.4}\times10^{14}$~M$_{\odot}$ falls at the lower end of these mass estimates but is consistent.

\section{SZ-Weak Lensing Scaling Relation}
\label{sec:szwl}

\begin{figure}
	\centering
	\includegraphics[width=1\columnwidth]{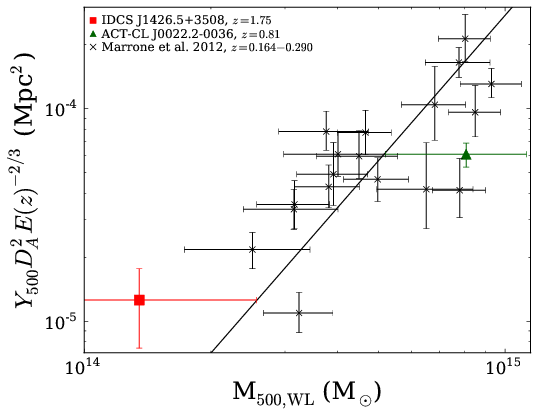}
	\caption{The SZ Comptonization parameter as a function of $M_{500}$ estimated from weak lensing. The black solid line is the $M_{500}-Y_{\mathrm{sph}}$ relation derived by \citet{marrone12}. Though at a significantly higher redshift than the \citet{marrone12} cluster sample, the IDCS~J1426.5+3508 SZ and weak lensing mass, plotted as the red square, are comparable to the \citet{marrone12} scaling relation. ACT-CL~J0022.2-0036 \citep{miyatake13} is also shown in green as an intermediate-redshift comparison.}
	\label{fig:scaling}
\end{figure}

The SZ Compton parameter $Y$ scales with mass as $Y\propto{M^{5/3}/({D_{A}^2E(z)^{-2/3}})}$ where $D_A$ is the angular diameter distance and $E(z)$ is the evolution of the Hubble parameter. We combine our weak lensing-derived mass estimate with the spherically averaged dimensionless Comptonization parameter ($Y_{\mathrm{sph},500}=7.9\pm3.2\times10^{-12}$) from \citet{brodwin12} to compare IDCS~J1426.5+3508 with the mass-$Y_\mathrm{sph}$ scaling relation derived by \citet{marrone12} using 18 Local Cluster Substructure Survey (LoCuSS) clusters in the range $z=[0.164,0.290]$. Figure~\ref{fig:scaling} shows our results alongside the \citet{marrone12} sample and scaling relation. In this figure, we match the $\Lambda$CDM cosmology assumed by \citet[][$\Omega_{\mathrm{M}}=0.27$, $\Omega_{\Lambda}=0.73$, $H_0 =73$~km~s$^{-1}$~Mpc$^{-1}$]{marrone12}. We also include the ACT-CL~J0022.2-0036 at $z=0.81$ \citep{miyatake13, reese12}, a higher redshift cluster with comparable SZ and weak lensing observations, for comparison. 

Our results for IDCS~J1426.5+3508 are statistically consistent with the low redshift data, indicating minimal redshift evolution in the \citet{marrone12} scaling relation. To quantify the redshift evolution, we fit a regression of the form
\begin{equation}
    \Big(\frac{1+z}{1+z_{\mathrm{M12}}}\Big)^{\alpha}=\Big(\frac{M_{500,\mathrm{WL}}}{10^{A}10^{14}M_{\odot}}\Big)\Big(\frac{10^{-5}\mathrm{Mpc}^{2}}{Y_{\mathrm{sph},500}D_{A}^2E(z)^{-2/3}}\Big)^{B}
\end{equation}
where $z_{\mathrm{M12}}=0.24$ is the average redshift of the LoCuSS sample and $A=0.367$ and $B=0.44$ are regression coefficients from \citet{marrone12}. Including the data for the $18$ LoCuSS clusters individually, ACT-CL~J0022.2-0036, and IDCS~J1426.5+3508, a least-squares fit yields $\alpha=-0.1\pm1.0$, within $1\sigma$ of zero. Thus, we find no significant evidence for evolution in the \citet{marrone12} scaling relation.

\section{Conclusions}
\label{sec:conclusions}

We have presented the weak lensing analysis of IDCS~J1426.5+3508. At $z=1.75$, this is the highest redshift galaxy cluster to be studied through weak lensing. We detect a tangential shear signal at a $2\sigma$-level significance. Assuming an NFW profile and a \citet{duffy08} mass-concentration relation, we fit for the cluster mass with a maximum-likelihood algorithm. Our mass estimate $M_{200,\mathrm{crit}}=2.3^{+2.1}_{-1.4}\times10^{14}$~M$_{\odot}$ is in agreement with estimates from X-ray, SZ, and strong lensing data. We also find that the SZ and weak lensing mass estimates agree well with the local scaling relation of \citet{marrone12}, with negligible evolution in the relation.

\citet{gonzalez12} initially discovered the strong-lensed arc behind IDCS~J1426.5+3508. If one makes standard assumptions for the galaxy cluster mass function, cosmological model, source galaxy redshift distribution, and cluster lensing cross section, then such an arc should not exist across the entire sky. The cluster lensing cross section increases with concentration, so one plausible explanation for this arc would be if the projected concentration of the cluster mass profile dramatically exceeds the typical value predicted for a cluster of this mass at this epoch. We are unable to determine a concentration as a free parameter for this cluster with the current data. However, a higher signal-to-noise weak lensing map and additional passbands to determine photometric redshifts would enable simultaneous fitting of the mass and concentration, improving Figure~\ref{fig:shear} and helping resolve the origin of the strong arc in this cluster.

\section*{Acknowledgments}
The authors thank the anonymous referee and Daniel Stern for their insightful suggestions and  Audrey Galametz for her help with the CANDELS-UDS data. Support for {\it HST} GO-program 11663, 12203, and 12994 was provided by NASA through a grant from the Space Telescope Science Institute, which is operated by the Association of Universities for Research in Astronomy, Inc., under NASA contract NAS 5-26555. We also acknowledge funding received from the National Science Foundation Graduate Research Fellowship under Grant No. DGE-1315138 (W.M.), NRF of Korea to CGER (M.J.J.), and Royal Society University Research Fellowship (R.M.). J.R. is supported by JPL, which is run by Caltech under a contract for NASA.

\bibliographystyle{apj}

\end{document}